# Prediction Capabilities of VLF/LF Emission as the Main Precursor of Earthquake

M.K.Kachakhidze[1] and N.K.Kachakhidze[1]

[1]{Saint Andrew the First-Called Georgian University}

## Abstract

Recent satellite and ground-based observations proved that in earthquake preparation period in the seismogenic area we have VLF/LF and ULF electromagnetic emissions.

According to the opinion of the authors of the present paper this phenomenon is more universal and reliable than other earthquake indicators. Hypothetically, in case of availability of adequate methodological grounds, in the nearest future, earth VLF/LF electromagnetic emission might be declared as the main precursor of earthquake. In particular, permanent monitoring of frequency spectrum of earth electromagnetic emission generated in the earthquake preparation period might turn out very useful with the view of prediction of large ($M \geq 5$) inland earthquakes.

The present paper offers a scheme of the methodology according to which the reality of the above given hypothesis can be checked up. To prove the prediction capabilities of earth electromagnetic emission we have used avalanche-like unstable model of fault formation and an analogous model of electromagnetic contour, synthesis of which, according to our opinion, is rather harmonious.

## 1 Introduction

Within the last decades many extremely interesting papers were dedicated to various geophysical phenomena revealed in earthquake preparation period and after earthquake periods, inclusive the study of earth EM emission in the period that preceded earthquake. Authors of the studies emphasize that genesis of all these phenomena (Changing of intensity of electro-telluric current in focal area; Perturbations of geomagnetic field in forms of irregular pulsations or regular short-period pulsations; Very low frequency (VLF) electromagnetic emission is observed on the earth surface; Perturbations of atmospheric electric field; Irregular changing of characteristic parameters of the lower ionosphere (plasma frequency, electron concentration, height of D-layer etc.); Irregular perturbations reach the upper ionosphere, namely F2-layer, for 2-3 days before the earthquake; There are appeared geomagnetic pulsations (0.02 – 0.1 Hz) in several hours or tenths of minutes before earthquakes. Intensity of electromagnetic emission increases in upper ionosphere in several hours or tenths of minutes before earthquake and sometimes it causes lighting) and their presence in definite period is associated with the earthquake preparation process, although, mechanisms of their origin can't be defined yet clearly.

It is known that earthquake prediction implies preliminary defining of the incoming earthquake place, time and magnitude, simultaneously.

Today, with this in view, we meet rather diverse and interesting papers, which are published in the world on the basis of ground-based and satellite data of earth EM emission observed in earthquake preparation period (Molchanov et al., 1992; Uyeda et all., 2000, Hattori, et al., 2004; Freund, et al., 2006;. Parrot, 2006; Biagi, et al., 2009;Eftaxias K., et al., 2009; Pulinets,2009; Rozhnoi,et al., 2009; Contadakis et,al., 2010; Papadopoulos, G. A., et al., 2010, Zlotnicki et al.,2010; Hayakawa, et al., 2013; Ouzounov, et al., 2013) If we had network of electromagnetic emission before earthquakes, we would be able, by means of satellite observations to differentiate the projection on the ground surface of the perturbed zone in the atmosphere-ionosphere boundary that approximately coincides with a zone of precursory activity.

This is evidenced by the results of studies connected with the earthquake of Italy occurring in 2009 (Rozhnoi et al., 2009). According to the opinion of the authors, in case of having of earth electromagnetic emission network, it is possible to determine 2-8 days prior to earthquake, by a definite precision, the place of an incoming earthquake.

It is also interesting that recorded anomalous magnetic field variations - the anomalous ULF signals were always detected in an area close to the epicenter about two weeks or several hours before earthquakes (Molchanov et al., 1992; Hayakawa et al., 1996; Hattori, et al., 2004; Liu et al., 2006; Kereselidze, et al., 2009; Chen et al, 2013).

Studying geodynamic processes the authors (Moroz et al., 2004), pay special attention to electromagnetic field. To characterize the field, they introduce the notion of telluric tensor the components of which depend on frequency of field changing, conductivity distribution in earth and orientation of coordinates axes.

It appeared that maximum value of telluric parameter was fixed on the direction the azimuth of which was 150 $^o$, that is, crosswise of seismofocal zone. At the same time, at the occurrence of large earthquakes (M=7.7 and M=6.9) their values decrease markedly. Decrease compared to background values begins approximately 1,5 months before earthquake. And after earthquake this parameter returns back to its value in approximately the same time (Moroz et al., 2004).

Some days before large earthquake, in the lower ionosphere there is confirmed an abnormal TEC variation over the epicenter (Ouzounov, et al., 2013).

Authors of the present paper have been explained the mechanism of earth EM emission in the period of earthquake preparation by analogous model of lithosphere-atmosphere-ionosphere (LAI) system's self-generated electromagnetic oscillations based on the classic electrodynamics (Kachakhidze et al., 2011). The physical analogy with the hypothetic ideal electromagnetic contour, the formation of which is assumed in epicenter area of incoming earthquake due to earth surface electric polarity changing is used. In this case the emphasis is made on the electromagnetic induction effect in local area of the space and not on the physical mechanism conditioning changing of earth potential value (atmospheric electric field tension inversion). The presence of such effect is proved in the papers (Bleier et al., 2009; Eftaxias et al., 2009).

Alongside with it, due to the fact that electromagnetic emission disturbance is conditioned by thermo-ionized channel (Kachakhidze, et al., 2011), in earthquake focus, it might well be that earth electromagnetic emission is of sector spreading, which, to a definite extent, refers to the main fault direction and epicenter area of incoming earthquake, which is also fixed by observations (Varotsos and Lazaridou, 1991; Sarlis et al., 1999; Uyeda et all., 2000; Moroz et al., 2004; Varotsos et all., 2006; Rosnoi, et al., 2009; Hayakawa, et al., 2013; Li, et al., 2013).

## 2 Discussion

Analogous model submitted by the authors of the present paper (Kachakhidze et al., 2011) is significant, since, on its base, by monitoring of electromagnetic emission existing in the period that precedes earthquake, it becomes possible to analyze the process of preparation and occurrence of large inland earthquake M≥5.

The above referred paper admits the formula, which connects with each other analytically the main frequency of the observed electromagnetic emission and the linear dimension (the length of the fault in the focus) of the emitted body:

$$\omega = \beta \frac{c}{l} \qquad (1)$$

where $\beta$ is the characteristic coefficient of geological medium and it approximately equals to 1. Of course it should be determined individually for each seismically active region, or for a local segment of lithosphere.

Comparing analogous and avalanche-like unstable models of fault formation (Mjachkin, et al., 1975), to determine reliability rate of their conformity to real process, we rely mainly on the data of earthquake which took place in Italy (L'Aquila) on 6 April, 2009 (Contadakis et,al., 2010; Papadopoulos, et al., 2010).

It is known that avalanche-like unstable model of fault formation is divided into three main stages (Mjachkin, 1975), (Fig.1): in case large earthquakes the first stage can go on for a dozen of months (Orihara,2012). At this stage chaotic formation of micro cracks without any orientation takes place.

This stage of formation of microcracks is reversible process - at this stage not only microcracks can be formed but also their the so-called "locked" can occur. Cracks created at this stage will be small (some dozen or hundred meter order) because the weak foreshock sequence may occur spatially distributed within the entire seismogenic area. For example such process was developed in case of earthquake of Italy in 2009: By the end of October 2008 the seismicity entered in the state of weak foreshock sequence which lasted up to the 26 March 2009. It is characteristic that the weak foreshock activity which developed from 28 October 2008 to 26 March 2009 spatially did not concentrated around the mainshock epicenter but it was widely distributed within the seismogenic area. The period when foreshock magnitude more or less increases and the so-called moderate foreshock occurs in region, is attributed to this stage too. This stage was fixed in case of L'Aquila earthquake: from 25 January 2009 to 26 March 2009, including this day. (Fig.2) (Papadopoulos, G. A., et al., 2010);

Such foreshocks can be called conditionally the "regional foreshocks" (Kachakhidze, M. et al., 2003). Because of short length of micro cracks and process reversibility first stage in the electromagnetic emission frequency range, according to our model (Kachakhidze et al., 2011) should be expressed by the discontinuous spectrum of MHz order emission (in radio diapason), which is proved by the latest special scientific literature (Eftaxias K., et al., 2009; Papadopoulos G., et al., 2010).

Thus, on the basis of analogous model, it can be stated that having of intermittent, high value MHz electromagnetic emission refers only to weak and moderate earthquakes (foreshocks), and it is not necessary for these foreshocks to be near the epicenter of the incoming main earthquake.

The second stage of the avalanche-like unstable model of fault formation is an irreversible avalanche process of already somewhat oriented microstructures, which is accompanied by inclusion of the earlier "locked" sections. Based on the analogous model, we have to suppose that this stage in the emission frequency spectrum should be expressed by MHz continuous spectrum already. Although, the values of electromagnetic emission frequency must gradually decrease. According to the avalanche-like unstable model, this process takes place few days (at about 10-14) before earthquakes, which is proved clearly by material of observations (Papadopoulos G., et al., 2010).

According to the avalanche-like unstable model, at the very stage gradual increase of cracks occurs (up to the kilometers order) at the expense of their uniting, to which, according to our model, from the formula (1) corresponds to the transition of MHz to kHz emission in the electromagnetic emission frequency spectrum.

If a rather large earthquake is prepared, of course, foreshock $M \geq 5$ is not excluded (as it was in case of L'Aquila earthquake) (Wu et al., 1978; Wang et al., 2006; Papadopoulos, G. A., et al., 2010). Because of this, electromagnetic spectrum can have VLF and LF frequency substitutions (Fig.3) (Papadopoulos, G. A., et al., 2010; Kachakhidze, M. et al, 2011).

At the final, third stage of the avalanche-like unstable model of fault formation the relatively big size faults use to unite into one - the main fault. This process, according to our model, in case of emission spectrum monitoring should correspond to gradual fall of frequencies in kHz, which according to the formula (1) refers to the increase of fault length in the focus.

Increase of crack length in focus refers to the increase of a magnitude of the incoming earthquake (Ulomov, 1993):

$$\lg l = 0.6M - 2.5 \qquad (2)$$

In case of L'Aquila earthquake, due to the fact that on 4.04.2009 the main frequency kHz was already fixed in the electromagnetic emission spectrum, (Papadopoulos, G. A., et al., 2010), the main fault in the earthquake focus should have been of kilometer order already.

Of course, association of cracks into one fault, which at the final stage of earthquake preparation proceeds intensely, will use definite part of energy accumulated in the focus and therefore, will result in its decrease. In such situation a period settles before a large earthquake (which can last from several hours to even 2 days), when in the focus a fault is already formed, while earthquake has not occurred yet, since accumulated tectonic stress is not yet sufficient to overcome the limit of strength of geological environment. The system, which is waiting for further "portion" of tectonic stress, is in the so-called "stupor", in the principle, the process of crack formation is not going on in it anymore, and respectively, electromagnetic emission would not take place.

Although, later, at the approach of critical value of tectonic stress, the balanced state in the system will be deranged and the earthquake will occur. This process is expressed correspondingly in the electromagnetic emission spectrum: some hours before the earthquake (up to 2 days) in the spectrum the emission intermittence is observed. Up to interruption of electromagnetic emission, by the use of final value of the main frequency of the spectrum (on the basis of the formula 1) we can determine, by a rather high accuracy, the length of expected fault of the future earthquake, that is, a magnitude of the incoming earthquake (Kachakhidze et al., 2011; Kachakhidze et al., 2012; Kachakhidze et al., 2013). We can expect renew of electromagnetic emission immediately before the earthquake.

In the period of electromagnetic emission monitoring the moment of interruption of emission spectrum is urgent for determination of time of occurrence of incoming earthquake, since at the final stage of earthquake preparation, very short time is needed to fill in the critical reserve of tectonic stress needed for main fault realization. It should be noted that this fact was experimentally proved for the case of L'Aquila earthquake. In this case this time lasted up to 1.5-2 days.

Often, in cases of rather large earthquakes we observe large foreshock too. The main shock can follow large foreshock rather swiftly. In this case, there is no reliable criterion, which can distinguish large foreshock from incoming earthquake. Example of this is L'Aquila earthquake, when it was considered that the incoming foreshock was the main shock.

This issue can be resolved by rather high accuracy on the basis of analogous model by EM emission monitoring. Namely, in case of M=5 earthquake, electromagnetic emission should equal to 95 kHz.

In this case fault length in the focus equals to approximately 3 km. (Kachakhidze et al., 2011; Kachakhidze et al., 2012). If after this earthquake electromagnetic emission still continues to exist and the frequency data still tend to decrease, it means that the process of fault formation in the focus of the earthquake is not completed yet and we have to wait for the main shock to occur.

Generally we should not expect stopping of electromagnetic emission after large earthquake but the frequency values in the spectrum must grow, which will refer to the fact that we should not expect the larger than the occurred earthquake but we have to wait for a series of aftershocks. This was well proved by the example of L'Aquila earthquake (Rozhnoi et al., 2009).

Since the processes of developing of foreshocks and aftershocks generally are connected with the fault formation process, it is clear that at this time too, VLF/LF electromagnetic emission will take place (Rozhnoi et al., 2009; Papadopoulos et al., 2010; Hayakawa et al., 2013). Analogous model on the basis of electromagnetic emission enable us to evaluate magnitude of each separate foreshock and aftershock.

Thus, good conformity of the above referred two models and capabilities of analogous model are evident on the example of real earthquake too.

Of course, in the whole world, rather interesting and significant works of geologists are dedicated to the study of issues of earthquake focus, foreshocks and aftershocks (Amato et al., 1998; Amato et al., 2008; Chiarabba et al., 2009), but we will not deal with it in this paper.

The present paper offers general, that is, "classical" picture of earthquake preparation and occurrence (foreshock – main earthquake – aftershock) on the basis of analogous model and avalanche-like unstable model of fault formation.

And finally, by the use of the formulas (1) and (2) for inland large M$\geq$5 earthquakes we can make the scale of dependence of incoming earthquake magnitude (even by the 0.1 accuracy) on the final, main frequency of electromagnetic emission fixed immediately before the earthquake.

Thus, monitoring of electromagnetic emission before the earthquake, on the basis of the offered models, enables us to follow, step by step, the process of earthquake preparation and make prognostic conclusions by definite precision.

As it is mentioned above, according to analogous model (Kachakhidze, et al., 2011), it is possible to detect the territory on the surface of the Earth in advance, where an earthquake is expected - the epicentral area of an incoming earthquake will be approximately limited to the territory where the earth surface will have positive potential towards atmosphere. Electromagnetic emission in kHz should take place namely on the territory adjoining the epicenter of a future large earthquake. Our opinions have been experimentally confirmed by the paper of K. Eftaxias (Eftaxias et al., 2009).

It should be stated that during electromagnetic VLF/LF and ULF emissions fixed before earthquake, there is a problem of differentiation of ground – based electromagnetic emission from magnetospheric emission, because of which it is impossible to prove reliably cause-and-effect relations among seismic and atmospheric (ionospheric) phenomena (Akhoondzadeh, et al., 2010; Zhang et al., 2010; Moldovan, et al., 2012; Masci, et al., 2013).

Of course, this problem is to be taken into consideration in the above mentioned researches.

## 3 Conclusions

EM emission that is considered as earthquake indicator is namely the main precursor, which "brings" for large inland (M$\geq$5) earthquake prediction the rich information about the stages of earthquake preparation process going on in the focus and in case of its permanent monitoring enables us to predict incoming earthquake by definite precision:

1. Appearance of intermittent electromagnetic emission spectrum in seismically active region, mainly in MHz range, refers to the fact that the process of large earthquake preparation has been started in the region; at this time, it is possible to fix the so-called "regional foreshocks" of relatively small magnitude and it is not excluded that this process will start some dozen months before the earthquake;

2. Due to the formation of significant size faults in focus, few days (approximately a fortnight) before earthquake, uninterrupted electromagnetic emission appears in MHz, kHz and ULF spectral range; In the spectrum initially MHz and ULF range emission should prevail, but periodically we should expect kHz range electromagnetic emission too;

3. On the next stage, in electromagnetic emission spectrum mainly kHz range frequencies dominate, which denotes that kilometer order main fault is in the process of formation in the focus already;

Shortly before the occurrence of earthquake, electromagnetic emission spectrum is only of kHz order, and it decreases swiftly;

In case of the devastating earthquakes (M$\geq$8.3) the main value of electromagnetic emission falls even to Hz order;

4. Few hours before the earthquake, or maximum 2 days before it, electromagnetic emission interrupts at all, which enables us to predict time of earthquake occurrence; At the moment of emission restarting, the earthquake occurs.

5. Value of the final main frequency of the spectrum emitted just before interruption of emission will enable us to define fault length in the earthquake focus, that is, the earthquake magnitude, by rather high precision;

6. Still more decrease of the main frequencies in electromagnetic spectrum after any large shock implies that the larger earthquake is expected and that the occurred shock was only a foreshock;

We can consider that a shock is the main earthquake if after it the main frequencies values in the emission spectrum begin to increase significantly. This effect is a prerequisite of starting of a series of aftershocks.

Thus, the analysis of electromagnetic emission spectrum enables us to differentiate clearly foreshocks and aftershocks from the main shocks;

7. The essential condition for determination of epicenter area of incoming earthquake, alongside with other possible methods, is that earth surface in this area should have positive potential permanently, for a rather long period (though for some weeks). Besides in addition to it, electromagnetic emission in kHz should take place namely on the territory adjoining the epicenter of an incoming large earthquake.

# References

Akhoondzadeh, M., Parrot, M., and Saradjian, M. R.: Electron and ion density variations before strong earthquakes (M > 6.0) using DEMETER and GPS data, Nat. Hazards Earth Syst. Sci., 10, 7–18, doi:10.5194/nhess-10-7-2010.

Amato, A., et al. (The 1997 Umbria-Marche, Italy, earthquake sequence: A first look at the main shocks and aftershocks, Geophys. Res. Lett., 25, 2861–2864, doi:10.1029/98GL5184, 1998.

Amato, A., and F. M. Mele. Performance of the INGV National Seismic Network from 1997 to 2007, Ann. Geophys., 51, 417– 431,2008.

Biagi, P. F., Castellana, L., Maggipinto, T., Loiacono, D., Schiavulli, L., Ligonzo, T., Fiore, M., Suciu, E., and Ermini, A.: A pre seismic radio anomaly revealed in the area where the Abruzzo earthquake (M = 6.3) occurred on 6 April 2009, Nat. Hazards Earth Syst. Sci., 9, 1551–1556, doi:10.5194/nhess-9- 1551-2009, 2009.

Bleier, T., Dunson, C., Maniscalco, M., Bryant, N., Bambery, R., and Freund, F.: Investigation of ULF magnetic pulsations, air conductivity changes, and infra red signatures associated with the 30 October Alum Rock M5.4 earthquake, Nat. Hazards Earth Syst. Sci., 9, 585–603, doi:10.5194/nhess-9-585-2009, 2009.

Chen, C. H., Hsu, H. L., Wen,S., Yeh, T. K., Chang, F. Y., Wang, C. H., Liu, J. Y., Sun, Y. Y., Hattori,K., Yen, H. Y. and Han,P.Evaluation of seismo-electric sing magnetic data in Taiwan anomalies using magnetic data in Taiwan. Nat. Hazards Earth Syst. Sci., 13, 597–604, doi:10.5194/nhess-13-597-201,2013.

Chiarabba,C. Amato,A., Anselmi,M., Baccheschi,P., Bianchi,I., Cattaneo,M., Cecere,G., Chiaraluce,L., Ciaccio,M.G., De Gori,P., . De Luca,G., Di Bona,M., Di Stefano,R., Faenza,L., Govoni,A., Improta,L., Lucente,F.P., Marchetti,A., Margheriti,L., Mele,F., Michelini,A., Monachesi,A.,Moretti,M., Pastori,M., Agostinetti, N. Piana., Piccinini,D., Roselli,P., Seccia,D., and Valoroso L. The 2009 L'Aquila (central Italy) MW6.3 earthquake: Main shock and aftershocks. GEOPHYSICAL RESEARCH LETTERS, VOL. 36, L18308, doi:10.1029/2009GL039627, 2009.

Contadakis, M. E., Biagi, P. F., and Hayakawa, M. (Eds.): Ground and satellite based observations during the time of the Abruzzo earthquake, Special Issue, Nat. Hazards Earth

Syst. Sci., issue102.html, 2010.

Eftaxias, K., Athanasopoulou, L., Balasis, G., Kalimeri, M., Nikolopoulos, S., Contoyiannis, Y., Kopanas, J., Antonopoulos, G., and Nomicos, C.: Unfolding the procedure of characterizing recorded ultra low frequency, kHZ and MHz electromagnetic anomalies prior to the L'Aquila earthquake as preseismic ones – Part 1, Nat. Hazards Earth Syst. Sci., 9, 1953– 1971, doi:10.5194/nhess-9-1953-2009, 2009.

Freund, F.T., A. Takeuchi and B.W.S. Lau. Electric currents streaming out of stressed igneous rocks – A step towards understanding pre-earthquake low frequency EM emissions, Phys. Chem. Earth, Parts A/B/C, 31, 389-396,2006.

Hattori, K.: ULF geomagnetic changes associated with large earthquakes, Terr. Atmos. Ocean. Sci., 15, 329–360, 2004;

Hayakawa, M., Kawate, R., Molchanov, O. A. and Yumoto, K., 1996, Results of ultra-lowfrequency magnetic field measurements during the Guam earthquake of 8 August 1993, Geophys. Res. Lett., 23, 241-244,1996;

Hayakawa, M., Hobara, Y., Rozhnoi, A., Solovieva, M., Ohta, K., Izutsu, J., akamura, T., Yasuda, Y., Yamaguchi, H., and Kasahara Y. The ionospheric precursor to the 2011 March 11 earthquake as based on the Japan-Pacific subionospheric VLF/LF network observation."Thales", in honour of Prof. Emeritus M.E. Contadakis, pp. 191-212, Ziti Publishing, Thessaloniki, 2013.

Kachakhidze, M., Kachakhidze, N., Kiladze, R.,Kukhianidze,V., Ramishvili, G. Relatively small earthquakes of Javakheti Highland as the precursors of large earthquakes occurring in the Caucasus. European Geophysical Union, Journal Natural Hazards and Earth System Sciences, v3., p.165-170 , 2003.

Kachakhidze, M. K., Kereselidze, Z. A., and Kachakhidze, N. K.The model of self-generated seismo-electromagnetic oscillations of the LAI system, Solid Earth, 2, 17–23, doi:10.5194/se-2-17-2011, 2011.

Kachakhidze, M., Kachakhidze,N.,Kereselidze,Z.,Ramishvili,G. and Kukhianidze,V. In connection with identification of VLF emission before L'Aquila earthquake. J "Natural Hazards and Earth System Science",12, 1009 – 1015, 2012.www.nat-hazards-earth-syst-sci.net/12/1009/2012/

Kachakhidze, M., Kereselidze,Z., Kachakhidze,N., Evaluation of Atmospheric Electric Field as Increasing Seismic Activity Indicator on the example of Caucasus Region. "Thales", in honour of Prof. Emeritus M.E. Contadakis, pp.213-225, Ziti Publishing, Thessaloniki, 2013

Kereselidze, Z., Kachakhidze,N., Kachakhidze, M., and Kirtskhalia,V.. Model of Geomagnetic Field Pulsations before Earthquakes Occurring. Georgian International Journal of  Science and Technology, Volume 2 Issue 2, pp. 167-178,  2009.

Li, M., Lu, J. , Parrot, M., Tan, H., Chang, Y., Zhang, X., and  Wang, Y. Review of unprecedented ULF electromagnetic anomalous emissions possibly related to the Wenchuan MS = 8.0 earthquake, on 12 May 2008. Nat. Hazards Earth Syst. Sci., 13, 279–286, doi:10.5194/nhess-13-279-2013;

Liu, J. Y., Chen, C. H., Chen, Y. I., Yen, H. Y., Hattori, K., and Yumoto, K.: Seismo-geomagnetic anomalies and M> 5.0 earthquakes observed in Taiwan during 1988–2001, Phys. Chem. Earth., 30, 215–222, 2006.

Masci, F., and De Luca,G. Some comments on the potential seismogenic origin of magnetic disturbances observed by Di Lorenzo et al. (2011) close to the time of the 6 April 2009 L'Aquila earthquake. Nat. Hazards Earth Syst. Sci., 13, 1313–1319, 2013

Mjachkin, V.I., Brace W.F., Sobolev G.A., Dieterich J.H., Two models for earthquake forerunners, Pageoph., vol. 113, Basel,1975.

Molchanov, O. A., Kopytenko, Yu. A., Voronov, P. M., Kopytenko, E. A., Matiashvili, T. G., Fraser-Smith, A. C., and Bernardi, A.: Results of ULF Magnetic field measurements near the epicenters of the Spitak (Ms 6.9) and Loma Prieta (Ms 7.1) earthquakes: comparative analysis, Geophys. Res. Lett., 19, 1495–1498, 1992.

Moldovan, I. A., Placinta, A. O., Constantin, A. P., Moldovan, A. S., and Ionescu, C.: Correlation of geomagnetic anomalies recorded at Muntele Rosu Seismic Observatory (Romania) with earthquake occurrence and solar magnetic storms, Ann. Geophys., 55,125–137, doi:10.4401/ag-5367, 2012.

Moroz, Y. F., Moroz, T. A., Nazareth, V. P., Nechaev, S. A., Smirnov. S. E. Electromagnetic field in studies of geodynamic processes. Complex seismological and geophysical researches of Kamchatka, (In Russian), 2004.

Orihara Yoshiaki , Kamogaw Masashi, Nagao Toshiyasu , and Uyeda Seiya (2012). Preseismic anomalous telluric current signals observed in Kozu-shima Island, Japan PNAS, November 20, , vol. 109 , no. 47 , 19035–19510, 2012.

Ouzounov,D., Pulinets,S., Romanov,A., Romanov,A.,Tsybulya,K., Davidenko,D., Kafatos,M. and Taylor,P. Atmosphere-Ionosphere Response to the M9 Tohoku EarthquakeRevealed by Joined Satellite and Ground Observations. Preliminary results. arXiv:1105.2841, 2013.

Papadopoulos, G. A., Charalampakis, M., Fokaefs, A., and Minadakis, G. Strong foreshock signal preceding the L'Aquila (Italy) earthquake (Mw6.3) of 6 April 2009, Nat. Hazards Earth Syst. Sci., 10, 19–24, doi:10.5194/nhess-10-19-2010, 2010.

Parrot, M. (Ed.) First results of the DEMETER micro-satellite, Planet. Space Sci., 54, 411-557, 2006.

Pulinets, S.A. (2009). Physical mechanism of the vertical electric field generation over active tectonic faults, Adv. Space Res., 44, 767-773, doi: 10.1016/j.asr.2009.04.038,2009.

Rozhnoi, A., Solovieva, M., Molchanov, O., Schwingenschuh, K., Boudjada, M., Biagi, P. F., Maggipinto, T., Castellana, L., Ermini, A., and Hayakawa, M.: Anomalies in VLF radio signals prior the Abruzzo earthquake (M = 6.3) on 6 April 2009, Nat. Hazards Earth Syst. Sci., 9, 1727–1732, doi:10.5194/nhess-9-1727-2009, 2009.

Sarlis, N., Lazaridou, M., Kapiris, P., and Varotsos, P.: Numerical model of the selectivity effect and the V/L criterion, Geophys. Res. Lett., 26, 3245–3248, 1999.

Ulomov, V. I.: Ordering of geostructure and seismicity in seismoactive regions. Seismisity and seismic zoning of Northern Eurasia, Moscow, Vol. 1, 27–31, 1993.

Uyeda, S., Nagao, T., Orihara, Y., Yamaguchi, T., and TakahashiI. Geoelectric potential changes: Possible precursors to earthquakes in Japan, Proc. Nat. Acad. Sci., 97, 4561–4566, 2000.

Varotsos, P. and Lazaridou, M.: Latest aspects of earthquake prediction in Greece based on seismic electric signals, Tectonophysics, 188, 321–347, 1991.

Varotsos, P., Sarlis, N., Skordas, E., and Lazaridou, M.: Additional evidence on some relationship between Seismic Electric Signals (SES) and earthquake focal mechanism, Tectonophysics, 412, 279–288, 2006.

Wang, K., Chen, Q.-F., Sun, S., and Wang, A., 2006, Predicting the 1975. Haicheng earthquake, Bull. Seismol. Soc. Am., 96, 757–795, doi:10.1785/0120050191, 2006.

Wu, K.-T., Yue, M.-S.,Wu, H.-Y., Chao, S.-L., Chen, H.-T., Huang, W.-Q., Tien, K.-Y., and Lu, S.-D.: Certain characteristics of the Haicheng earthquake (M=7.3) sequence, Chinese Geophysics,1,1978.

Zhang, X., Shen, X., Liu, J., and Ouyang, X.: Ionospheric perturbations of electron density before theWenchuan Earthquake, Int. J. Remote Sens, 31, 3559–3569, 2010.

Zlotnicki, J., F. Li and M. Parrot. Signals recorded by DEMETER satellite over active volcanoes during the period 2004 August – 2007 December, Geophys. J.2010.

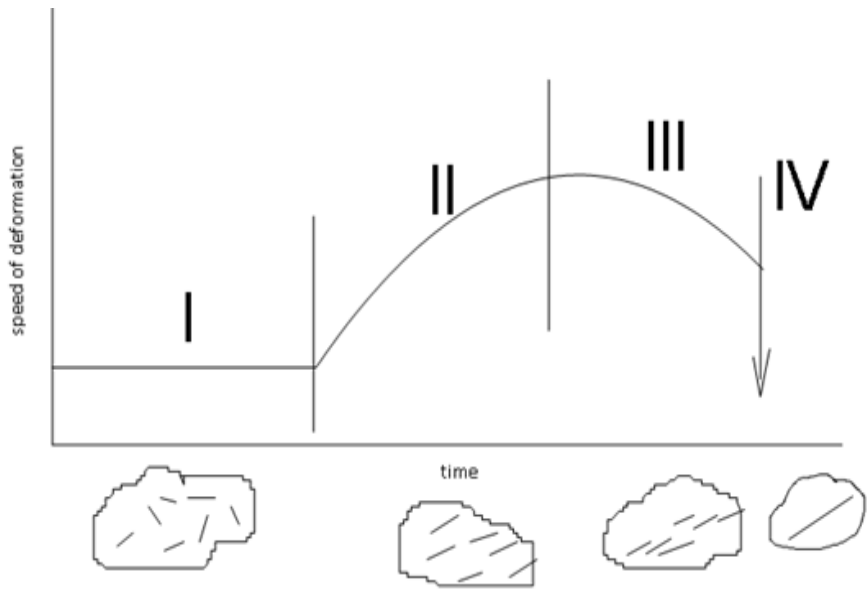

Figure 1. Scheme of fall-unstable model of fracture origination

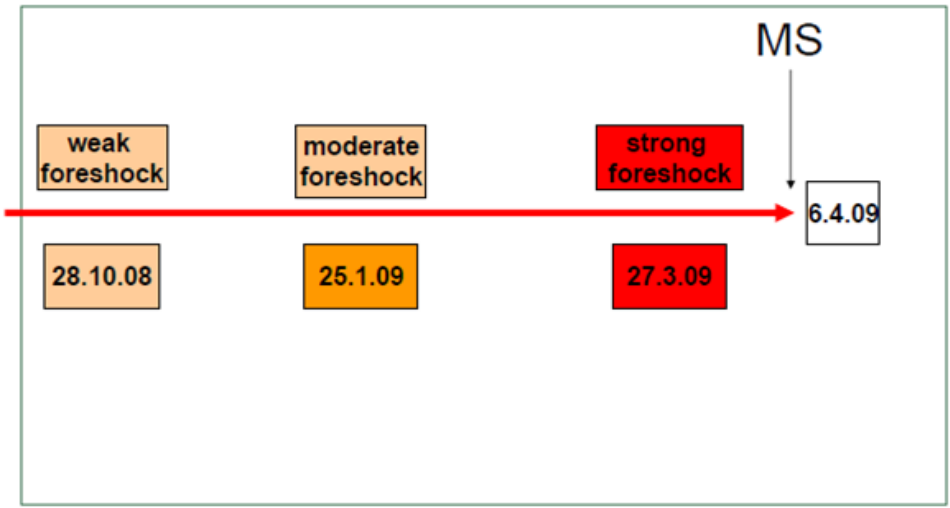

Figure 2. Evolution of foreshock activity

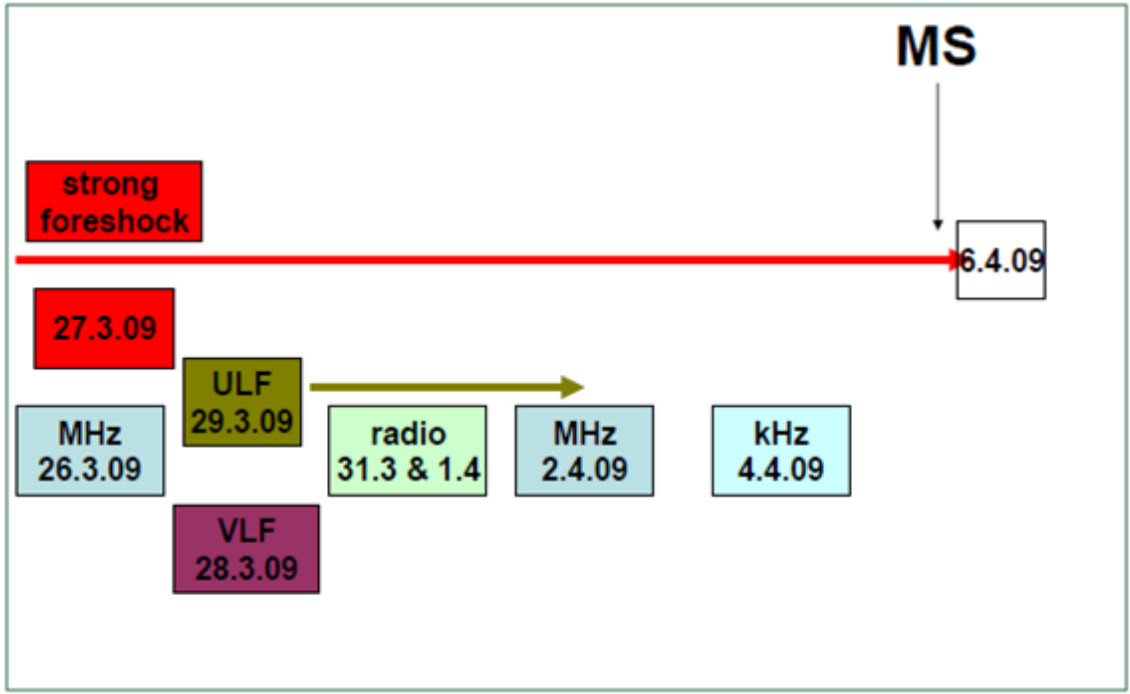

Figure 3. Evolution of EM emission